\documentclass[]{fairmeta}
\usepackage{pifont}
\usepackage{pgfplots}
\pgfplotsset{compat=1.18}
\usepackage[ruled,vlined]{algorithm2e}

\usepackage[utf8]{inputenc}
\usepackage{enumitem}
\usepackage{amsmath}
\usepackage{amssymb}
\definecolor{C0}{rgb}{0.1216, 0.4667, 0.7059}
\definecolor{C1}{rgb}{1.0000, 0.4980, 0.0549}
\definecolor{C2}{rgb}{0.1725, 0.6275, 0.1725}
\definecolor{C3}{rgb}{0.8392, 0.1529, 0.1569}
\definecolor{C4}{rgb}{0.5804, 0.4039, 0.7412}

\title{Improving ML Attacks on LWE with Data Repetition and Stepwise Regression}

\author[1,*]{Alberto Alfarano}
\author[2]{Eshika Saxena}
\author[3]{Emily Wenger}
\author[1,*, \dagger]{Fran\c{c}ois Charton}
\author[2,\dagger]{Kristin Lauter}

\affiliation[1]{Axiom Math}
\affiliation[2]{FAIR at Meta}
\affiliation[3]{Duke University}

\contribution[*]{Work done at Meta}
\contribution[\dagger]{Joint last author}

\abstract{The Learning with Errors (LWE) problem is a hard math problem in lattice-based cryptography.  In the simplest case of binary secrets, it is the subset sum problem, with error.
Effective ML attacks on LWE were demonstrated in the case of binary, ternary, and small secrets, succeeding on fairly sparse secrets. 
The ML attacks recover secrets with up to 3 active bits in the ``cruel region'' ~\citep{coolcruel} on samples pre-processed with BKZ. We show that using larger training sets and repeated examples enables recovery of denser secrets. Empirically, we observe a power-law relationship between model-based attempts to recover the secrets, dataset size, and repeated examples. We introduce a stepwise regression technique to recover the ``cool bits'' of the secret.}

\date{\today}
\correspondence{Kristin Lauter at \email{klauter@meta.com}}

\begin{document}

\maketitle

\section{Introduction}

Most current public-key cryptosystems, used to secure many online interactions, are susceptible to attacks based on Shor's algorithm~\citep{shor1994}, a fast method for integer factorization on a quantum computer.  To counter this, a new class of systems, known as post-quantum cryptography (PQC), have been developed and are currently being standardized~\citep{nist2022finalists}. Assessing their potential weaknesses and limitations is an active field of research.

Many PQC systems are based on a hard math problem known as Learning With Errors (LWE)~\citep{Reg05}. The security of LWE stems from the hardness of recovering a secret vector of integers modulo $q$ from its noisy dot products with random vectors: find $\mathbf{s} \in \{0, \dots, q-1\}^n$, given $m$ pairs $(\mathbf a_i, b_i=\mathbf a_i \cdot \mathbf s + \epsilon_i \mod q)_{i\in \{1\dots m \}}$, with $\mathbf a_i$ randomly sampled from a uniform distribution over $\{0,\dots, q-1\}^n$, and $\epsilon_i$ sampled from a discrete Gaussian distribution with small standard deviation, width $3$ or $3.2$. 

LWE is hard when $\mathbf{s}$ is chosen uniformly modulo $q$, for large values of $n$, and not too large moduli $q$. Alternatives with \emph{small} secrets,  (i.e. binary, ternary or small integer secrets, where non-zero coordinates are $1$, $1$ or $-1$, or small integers, respectively), and \emph{sparse} secrets (few non-zero entries), are attractive for efficiency of homomorphic encryption.  
The potential weaknesses of these variants, for practical values of $n$ and $q$, is under debate.
This is the subject of this paper.

SALSA, a transformer-based ML attack on LWE with small and sparse secrets, was proposed in prior work~\citep{salsa, picante, verde, fresca}. SALSA uses LWE pairs sharing the same secret, $(\textbf{a}, b=\textbf{a}\cdot \textbf{s}+\epsilon \mod q)$, to train a transformer to predict $b$ from $\mathbf a$. Once the model learns to predict better than randomly, secret coordinates can be recovered by comparing model predictions for related values of $\mathbf a$. Later versions of SALSA~\citep{picante, verde, fresca} introduced a pre-processing step: LWE pairs, sub-sampled from a small number of eavesdropped samples, are first reduced by lattice reduction algorithms (BKZ~\citep{BKZ, BKZ2}) and then used to train the transformer. BKZ reduction starkly reduces the later coordinates of $\mathbf a$ (the cool region), while leaving the $c$ first coordinates (the cruel region) relatively unreduced~\citep{coolcruel}. These papers show that preprocessing with BKZ reduction enables the recovery of secrets with larger Hamming weight $h$ (number of non-zero secret bits)~\citep{picante, verde, fresca}.

However, ML attacks on LWE are constrained by the model's ability to learn the modular dot product. In particular, models tend to struggle when the dot product $\mathbf a \cdot \mathbf s$ grows large~\citep{verde}, and potentially wraps around the modulus many times. This happens for example when there are more than three non-zero bits in the unreduced (cruel) region of the secret~\citep{benchmarking}. This limits the success of ML attacks as the Hamming weights of secrets increases. For example, for $n=512$, with $\log_2 q=28$, after BKZ reduction, the cruel region in~\cite{coolcruel} is the first $228$ bits. Therefore, most secrets with $h\geq 10$ will have $4$ cruel bits or more, and thus were not recovered in past work.

In this paper, we explore methods that allow ML attacks to recover secrets with more than $3$ cruel bits. We explore the impact of larger training sets and repeated examples. We also introduced two new methods for recovering cool bits, based on stepwise regression~\citep{efroymson1960multiple}. Overall, \textbf{our contributions} are:

\begin{itemize}[nosep,left=0pt]
    \item training on \emph{larger datasets} and \emph{repeated} examples, which allows us to  recover binary and ternary secrets with up to \emph{$8$ cruel bits},
    \item replacing the linear regression algorithm used to recover cool bits~\citep{benchmarking} with \emph{stepwise regression},
    \item generating 400 million \emph{synthetic data samples} to explore secret recovery as a function of training set size, repetition, secret Hamming weight and number of cruel bits, in four LWE settings,
    \item defining an empirical \emph{scaling law} relating secret recovery to amount of training data and data repetition.
\end{itemize}

Overall, our attack recovers secrets with larger Hamming weights than prior work (Table~\ref{tab:main_results}), and demonstrates that the limitation of ML attacks to secrets with $3$ cruel bits can be overcome.

\begin{table}[h]
    \centering
    \small
    \caption{{\bf Largest Hamming weights recovered} compared to C\&C: Cool and Cruel attack~\citep{coolcruel} and SALSA: VERDE/PICANTE~\citep{verde,picante}.}
    \label{tab:main_results}

    \begin{minipage}[t]{0.48\textwidth}
        \centering
        \begin{tabular}{lccc}
            \toprule
            Settings & Ours & Prior ML & C\&C \\
            \midrule
            n=256 $\log_2 q=12$ & \textbf{14} &  8 & 12 \\
            n=256 $\log_2 q=20$ & \textbf{70} & 33 & - \\
            n=512 $\log_2 q=28$ & \textbf{12} & - & \textbf{12} \\
            n=512 $\log_2 q=41$ & \textbf{75} & 63 & 60 \\
            \bottomrule
        \end{tabular}
        \captionsetup{justification=centering}
        \caption*{Binary secrets}
    \end{minipage}
    \hfill
    \begin{minipage}[t]{0.48\textwidth}
        \centering
        \begin{tabular}{lccc}
            \toprule
            Settings & Ours & Prior ML & C\&C \\
            \midrule
            n=256 $\log_2 q=12$ & \textbf{12} & 9 & - \\
            n=256 $\log_2 q=20$ & \textbf{55} & 24 & - \\
            n=512 $\log_2 q=28$ & \textbf{10} & - & - \\
            n=512 $\log_2 q=41$ & \textbf{75} & 66 & - \\
            \bottomrule
        \end{tabular}
        \captionsetup{justification=centering}
        \caption*{Ternary secrets}
    \end{minipage}
\end{table}

\section{Related Work}

SALSA~\citep{salsa} was the first ML attack demonstrated on LWE. It uses a shallow sequence-to-sequence transformer~\citep{transformer17}, with shared layers~\citep{dehghani2018universal}, trained on $2$ million unreduced LWE pairs, and recovers secrets from the trained model with a basic distinguisher. Limited to dimensions up to $128$ and Hamming weight $3$, it was a proof of concept: all secrets recovered by SALSA could be found with exhaustive search. It also required millions of eavesdropped LWE pairs with the same secret, an unrealistic assumption. 

PICANTE~\citep{picante} introduced pre-processing. It only requires $4n$ eavesdropped LWE pairs (a realistic assumption) which are sampled to form $n \times n$ matrices $\mathbf A$, and reduced by BKZ to produce matrices $\mathbf{RA}$ with a low standard deviation of their entries. Applying the same transformation to $\mathbf b=\mathbf A\cdot \mathbf s + \epsilon$ yields reduced LWE pairs with the same secret, but increased noise. A  transformer is then trained from $4$ million reduced pairs,  which can recover sparse binary secrets with $h=31$ for $n=256$ and $\log_2 q=23$, and $h=60$ for $n=350$ and $\log_2 q=32$. 

VERDE~\citep{verde} improved reduction techniques and distinguishers, recovering secrets with $h=63$ for $n=512$ and $\log_2 q=41$. VERDE  extended the recovery mechanism to ternary and small secrets, demonstrating that ternary secrets are not more secure than binary. VERDE also provided a theoretical explanation of the role of reduction: models struggle to learn modular additions that exceed a multiple of the modulus. FRESCA~\citep{fresca} introduced an encoder-only model and angular embeddings, an architecture for modular arithmetic which halves the length of inputs, allowing secret recovery for $n=1024$ and $\log_2 q=50$.

The difficulty of learning modular addition, first observed by~\citep{palamasinvestigating}, was further studied by~\citep{saxena2025making}. They showed that noiseless modular addition of many integers can be learned by adding a regularizer to the loss introduced in FRESCA and adding sparser examples to the training data, in a manner of curriculum learning.

\citep{coolcruel} offer a different perspective on BKZ reduction. They observe that the reduction of $\mathbf a$ is concentrated to the last coordinates (the cool region), while  the $c$ first (the cruel region) stay unreduced. Due to the reduction, the cool bits of the secret are easy to recover once the cruel bits are known. Therefore, recovering secrets from reduced LWE pairs amounts to discovering their cruel bits, and ML attacks only recover secrets with up to three cruel bits. The paper also proposed a non-ML attack, where all possible cruel bits are enumerated, and the cool bits are then guessed. It performs well for small dimensions and reduced data~\citep{benchmarking}, but scales exponentially with the Hamming weight, and requires a large number of LWE pairs for cool bit recovery. \citep{benchmarking} provides a comparison between SALSA, Cool \& Cruel, and classic attacks on LWE for $n=256, 512, \text{and } 1024$.

Training from repeated examples was proposed by~\citep{charton2024emergentpropertiesrepeatedexamples}. They show that multiplication modulo $67$, a task that transformers cannot usually learn when trained from large datasets of single-use examples, can be learned with $100\%$ accuracy when models are trained from smaller sets of repeated data.~\citep{saxena2025making} demonstrate a similar effect for modular addition. These papers are the original inspiration for the methods presented in section~\ref{sec:repetition}.

\section{Experimental settings}\label{sec:settings}

\textbf{LWE settings.} We consider four sets of parameters for LWE problems (see Table~\ref{tab:red_params} in Appendix~\ref{app:bkz}): $(n=256, \log_2 q=12)$, $(n=512, \log_2 q=28)$, $(n=256, \log_2 q=20)$, and $(n=512, \log_2 q=41)$. In the first two settings, the small modulus size makes BKZ reduction less effective, and the size of the cruel region is large. The last two settings allow for more reduction, smaller cruel region, higher recoverable Hamming weights, but a lot of cool bits. In all four settings, the standard deviation of error is $3$. These settings are also used in prior work ~\citep{verde, coolcruel}.

We choose $n$ for which running lattice BKZ reduction is possible with reasonable resources but running brute force attacks is not feasible. For $n=256$, recovering a $h=14$ secret takes $10^{22}$ attempts, each requiring thousands of operations. This would take several months on the fastest current supercomputer ($10^{18}$ FLOPS), at the frontier of current brute force capability. For $n=512$, recovering a $75$-bit secret requires $10^{91}$ attempts, which is far beyond any brute-force attack capability.

\textbf{Data generation and reduction.} We implement the AI-based attack described in Fresca~\citep{fresca, benchmarking}. 
Starting from $4n$ random uniform vectors $\textbf a_i\in \mathbb Z_q^n$, we sample $m\leq n$ vectors without replacement ($m=0.875n$ usually),
and stack them in a matrix $\mathbf{A} \in \mathbb{Z}_q^{m \times n}$. We construct
$\mathbf{\Lambda} =
\left[\begin{smallmatrix}
q\cdot \mathbf{I}_m & \mathbf{0} \\
\mathbf{A} & \omega\cdot \mathbf{I}_n
\end{smallmatrix}\right]$$\in \mathbb{Z}^{(m+n) \times (m+n)}$.
The matrix $\mathbf \Lambda$ is then reduced, using  BKZ2.0 interleaved with flatter~\citep{coolcruel}, yielding a $(n+m)\times n$ matrix
which has entries with a smaller variance. For a given secret $\mathbf s$, the LWE pairs $(\textbf A, \textbf b = \mathbf{A}\cdot \mathbf s + \mathbf{\epsilon})$ are then transformed by the matrix $\mathbf R$ extracted from the reduction process to  provide $n+m$ reduced LWE pairs $(\textbf {RA}, \textbf {Rb})$, with the same secret, but a larger error $\epsilon$ (the parameter $\omega$, set to $10$~\citep{picante} controls the trade-off between reduction and error amplification). This process is repeated to generate the train and test sets. We utilized CPUs with 750 GB of RAM and around 42M CPU hours to generate the BKZ-reduced datasets, please refer to Appendix~\ref{app:bkz} for a more detailed breakdown.

After reduction, the variance of the last $n-c$ columns of $\mathbf{RA}$ is reduced to less than half the standard deviation of an uniform distribution, $\frac q {\sqrt{12}}$. The first $c$ columns, the cruel region, stay unreduced~\citep{coolcruel}. For a given $n$, the size of cruel region $c$ decreases as the modulus $q$ increases. Table~\ref{tab:red_params} provides the reduction parameters for the four settings used in this paper.

\textbf{Synthetic data.} BKZ reduction uses a large amount of resources: for $n=512$, it takes about one minute on one CPU to produce one reduced-LWE pair. For our experiments with large training sets (Sections~\ref{sec:repetition}), we created $400$ million reduced pairs for $n=256$ and $\log_2 q= 20$. For the other settings, we only created between $4$ and $60$ million reduced pairs using BKZ reduction. We also generated, for all setting, $400$ million synthetic reduced vectors $\mathbf a$, with the same coordinate variance as the reduced pairs (i.e. $c$ unreduced coordinates, and $n-c$ reduced with the same amount of reduction as BKZ reduction). The results from section~\ref{sec:results} indicate that secret recovery rates are the same for models trained on BKZ-reduced and synthetic data. An additional discussion can be found in  Appendix~\ref{app:synthetic}. We utilized around 1,000 CPU hours to generate the synthetic datasets.

\textbf{Model architecture.} The reduced data is used to train an encoder-only transformer with $4$ layers, embedding dimension $d=256$ (for settings with the larger values of $q$) and $d=512$ (for the harder settings with smaller $q$) and a ratio of dimension to heads of $d/h=64$. Each coordinate $a_i$ of the input vector $\mathbf a$ is encoded by the angular embedding introduced in ~\citep{fresca}, as the point $(\cos(\frac{2\pi a_i}{q}), \sin(\frac{2\pi a_i}{q})) \in \mathbb  R^2$ and mapped onto $d$-dimensional space by a learnable linear layer. Two learnable positional embeddings are added to the input: an absolute position embedding (from $1$ to $n$), and a binary embedding indicating whether the current position is in the cruel or cool region. Overall, the input vector for coordinate $a_{i}$ of $\mathbf{a}$ is $\mathrm{Emb(a)} = \left( \mathrm{Emb}_1(a_{i})+\mathrm{Emb}_2(i)+\mathrm{Emb}_3(1_{i\leq c}) \right)$, with $\mathrm{Emb}_1$, $\mathrm{Emb}_2$, $\mathrm{Emb}_3$ $2\times d$, $n\times d$ and $2\times d$ matrices. $\mathrm{Emb}_3$, the cool and cruel embedding, is an improvement we introduce which allows for better cruel bit recovery rates over previous works. For $n=256$, $\log_2 q=20$, with $400$ million samples, it boosts the recovery rate of secrets with $4$ cruel bits from $2/5$ (without the embedding) to $5/5$ (see Appendix~\ref{app:cc_emb_ablation} for additional details).

As in FRESCA~\citep{fresca}, the transformer output, a sequence of $n$ vectors in $\mathbb R^d$, is max-pooled, and processed by a linear layer of size $d\times 2$. Its output $(x,y)$ is decoded as the integer $p$ such that $(\cos(\frac{2\pi p}{q}), \sin(\frac{2\pi p}{q}))$ is closest to $(x,y)$.

\textbf{Model training.} All experiments run on $1$ V100 GPU with $32$ GB of memory, for $2$ billion examples at most. Training time is around 5 days. The model is trained to minimize the mean-square error (MSE) between model predictions and correct answers, using the Adam optimizer~\citep{kingma2014adam} with mini-batches of $256$. In theory, all model predictions should lie on the unit circle, but prior work~\citep{saxena2025making} observed that they tend to drift towards the origin at the beginning of training. They proposed to add the penalty  $\alpha \left(r^2+\frac 1 {r^2}\right)$, with $r^2=x^2+y^2$ and $\alpha=0.1$, to the MSE loss. At the beginning of training, when the model makes random predictions of $b$, the MSE loss has a local minimum at the origin. 
Unfortunately, the loss at $(0,0)$ is independent of $b$, therefore the closer predictions are to the origin, the less the model learns. To prevent such a collapse, we consider a more general penalty of the form $P(r)=\alpha r^2 + \beta / r^2$ to the MSE loss. For the remaining sections of the paper, we set $\alpha=\beta=0.1$, and we defer the discussion on the best parameters with an ablation study across different settings to Appendix~\ref{app:collapse}.

\textbf{Secret recovery.} At the end of each epoch (usually after $2.5$ million training examples), an attempt is made to recover the secret. The distinguisher introduced in~\cite{verde} is run on $1000$ reduced LWE samples and ranks the $c$ cruel columns of the secret by their likelihood of being different from zero. The rank is then used to generate $15,000$ model based attempts of the cruel bits. For each of them, the cool bits are then estimated using stepwise regression, see Section~\ref{sec:stepwise}, producing a secret guess that is evaluated using the statistical test defined in PICANTE~\cite{picante}[Section 4.3].
If the correct secret is discovered, the process stops, else another training epoch is run.

\section{Recovering higher Hamming weights with large sets of repeated examples}\label{sec:repetition}

Prior work suggested that ML attacks on LWE samples do not recover secrets with $h>3$ when trained on non-reduced data, or more than $3$ cruel bits when trained  on reduced data ~\citep{coolcruel,benchmarking}. In this section, we demonstrate that larger training sets and repeated examples allow to recover more than $3$ unreduced secret bits.

We first consider models trained on \textbf{non-reduced data}, as in the original SALSA paper. This setting is the cleanest possible, as there can be no side effects due to BKZ reduction, cool bits, or noise amplification. We experiment with $n=64$ and $\log_2 q=20$, and secrets with $3\leq h \leq 6$, in the presence and absence of noise. For each value of $h$, we train models on $3$ different secrets, and report success if one is recovered at least.

\begin{table}[h!]
    \small
    \centering
    \caption{Secret recovery for different Data Budgets (DB), and repetition levels.  --: no secret recovered, \ding{51}: recovery in the noiseless case only, \ding{51}\ding{51}: recovery in all cases.}
    \label{tab:toy}
    \begin{tabular}{ll|cccc}
        \toprule
        Hamming & Repetition & \multicolumn{4}{c}{Data budget} \\
        weight & level & 1M & 4M & 20M & 50M \\
        \midrule
        \multirow{5}{*}{4} 
        & 1x  & -- & \ding{51} & \ding{51} & \ding{51}\ding{51} \\
        & 2x  & -- & \ding{51} & \ding{51} & \ding{51}\ding{51} \\
        & 4x  & -- & \ding{51}\ding{51} & \ding{51}\ding{51} & \ding{51}\ding{51} \\
        & 10x & \ding{51} & \ding{51}\ding{51} & \ding{51}\ding{51} & \ding{51}\ding{51} \\
        & 20x & \ding{51} & \ding{51}\ding{51} & \ding{51}\ding{51} & \ding{51}\ding{51} \\
        \midrule
        \multirow{5}{*}{5} 
        & 1x  & -- & -- & -- & -- \\
        & 2x  & -- & -- & -- & -- \\
        & 4x  & -- & -- & \ding{51} & \ding{51} \\
        & 10x & -- & \ding{51} & \ding{51} & \ding{51}\ding{51} \\
        & 20x & -- & \ding{51} & \ding{51}\ding{51} & \ding{51}\ding{51} \\
        \midrule
        \multirow{5}{*}{6} 
        & 1x  & -- & -- & -- & -- \\
        & 2x  & -- & -- & -- & -- \\
        & 4x  & -- & -- & -- & \ding{51} \\
        & 10x & -- & -- & \ding{51} & \ding{51} \\
        & 20x & -- & -- & \ding{51} & \ding{51}\ding{51} \\
        \bottomrule
    \end{tabular}
\end{table}

As in SALSA, secrets with Hamming weight $3$ are always recovered, even when the model is trained without repetition on one million sample only. Table~\ref{tab:toy} presents our findings for $h=4$ to $6$. Secrets with $h=4$ are recovered without repetition for large data budgets, but repetition enables recovery even for small data budgets. Recovery of secrets with larger Hamming weights requires large data budgets \emph{and} repetition. These results, a major improvement over SALSA, demonstrate the potential benefit of large training sets of repeated examples.

Next we experiment with \textbf{reduced data}, for $n=256$, $\log_2 q= 20$ (cruel region size $34$, Table~\ref{tab:re1}), and $n=512$, $\log_2 q=28$ (cruel region size $228$, Table~\ref{tab:re2}). We consider secrets with $h=4$ and $5$. For each setting and value of $h$, we sample $4$ different secrets, and train $16$ models (with different weight initializations). An experiment is successful if one model recovers all cruel bits, and we report the number of secrets recovered out of four. Notation ``$X|Y$'' indicates $X$ recoveries with $4$ cruel bits and $Y$ recoveries with $5$. 

For $n=256$, secrets with $4$ cruel bits are recovered, given very large data budgets ($400$ million different examples). Repetition allows recovery from smaller datasets. Secrets with $h=5$ can only be recovered with large training sets. For dimension $512$, secrets with $4$ cruel bits can be recovered from $20$ million of examples, repeated $10$ times. Secrets with $5$ cruel bits can be recovered with a training budget of $200$ million examples. Models trained on large sets of repeated examples do recover secrets with more than $3$ cruel bits, a clear improvement over previous ML attacks.

\begin{table}[!h]
\centering
\small
    \begin{minipage}[t]{0.48\textwidth}
        \centering
        \caption{\bf $\boldsymbol{n=256, \log_2 q=20}$}
        \label{tab:re1}
        \begin{tabular}{l|cccccc}
            \toprule
             Data & \multicolumn{6}{|c}{4$\vert$5 cruel bits} \\
            budget & 1x & 2x & 5x & 10x & 20x & 100x\\
            \midrule
            20M   & 0$\vert$0  & 0$\vert$0  & 0$\vert$0  & 0$\vert$0   & 0$\vert$0   & 0$\vert$0    \\
            50M   & 0$\vert$0  & 0$\vert$0  & 0$\vert$0  & 0$\vert$0   & 1$\vert$0   & -      \\
            100M  & 0$\vert$0  & 0$\vert$0  & 1$\vert$0  & 1$\vert$0   & 1$\vert$0   & -      \\
            200M  & 0$\vert$0  & 1$\vert$0  & 2$\vert$0  & 3$\vert$1   & -     & -      \\
            400M  & 3$\vert$1  & 4$\vert$1  & 4$\vert$1  & -     & -     & -     \\
           \bottomrule
        \end{tabular}
    \end{minipage}
    \hfill
    \begin{minipage}[t]{0.48\textwidth}
        \centering
        \caption{\bf $\boldsymbol{n=512, \log_2 q=28}$}
        \label{tab:re2}
        \begin{tabular}{l|cccccc}
            \toprule
            Data & \multicolumn{6}{|c}{4$\vert$5 cruel bits}\\
            budget & 1x & 2x & 5x & 10x & 20x & 100x \\
            \midrule
            20M   & 0$\vert$0  & 0$\vert$0  & 0$\vert$0  & 1$\vert$0   & 1$\vert$0   & 1$\vert$0    \\
            50M   & 0$\vert$0  & 1$\vert$0  & 1$\vert$0  & 1$\vert$0   & 1$\vert$0   & -    \\
            100M  & 0$\vert$0  & 1$\vert$0  & 1$\vert$0  & 1$\vert$0   & 2$\vert$0   & -    \\
            200M  & 0$\vert$0  & 2$\vert$0  & 2$\vert$0  & 2$\vert$1   & -     & -    \\
            400M  & 1$\vert$0  & 4$\vert$0  & 4$\vert$1  & -   & -     & -    \\
           \bottomrule
        \end{tabular}
    \end{minipage}

    \vspace{0.5em}
    \captionsetup{justification=centering}
    \caption*{Number of secrets recovered. ``-'' indicates experiments could not be run: compute budget is too large}
\end{table}

These results also shed new light on the role of cool bits. In theory, secrets should be easier to recover with $n=256$, $\log_2 q=20$ than with $n=512$, $\log_2 q=28$: the dimension is smaller, the the BKZ-reduction rate is much better ($34$ cruel columns vs $228$). Yet, with repetition, secrets with $4$ cruel bits  can be recovered with $20$ million examples only for $n=512$, vs $50$ million for $n=256$. Somehow, a higher reduction rate, and more cool bits, seems to complicate the recovery of cruel bits. In the next section, we investigate this counter-intuitive observation.

\section{Taming the cool bit noise: stepwise regression}\label{sec:stepwise}

The ML-based SALSA attacks recover secrets by comparing model predictions for related values of $\mathbf{a}$. For example for binary secrets, if $\mathbf a'=\mathbf {a}+ \frac q 2 \mathbf e_i$, with $\mathbf e_i$ the $i$-th standard basis vector, then the difference between the associated values of $b = \mathbf a \cdot \mathbf s + \epsilon$:  $$b'-b=(\mathbf a' - \mathbf a) \cdot \mathbf s + \epsilon' - \epsilon=\frac q 2 \mathbf s_i+ \epsilon' - \epsilon \mod q$$ has mean $0$ if $\mathbf s_i=0$, and $q/2$ if $\mathbf s_i=1$. If the transformer $\mathcal T$ produces good predictions of $b$ and $b'$, i.e. $\mathcal T(\mathbf a)\approx b$ and  $\mathcal T(\mathbf a')\approx b'$, then the difference $|\mathcal{T}(\mathbf a)-\mathcal{T}(\mathbf a')|$, averaged over vectors $\mathbf a$, reveals the corresponding secret bit.

Previous research (\citep{verde}[Section $6$]) suggests that transformers struggle to learn modular dot products when the sum $|\mathbf a \cdot \mathbf s|$ becomes larger than a multiple of $q$. With reduced data, this can happen in two cases: when the number of (unreduced) cruel bits in the secret exceeds a relatively small value, \textbf{but also} when the size of the reduced, cool, region becomes large.
We believe this accounts for the observation, at the end of the previous section, that the cruel bits for the ``hard setting'' $n=512$ $\log_2 q=28$, were easier to recover, than those of the easier setting $n=256$ $\log_2 q=20$, which enjoyed a higher reduction factor. In this section, we investigate cool bit recovery, especially in the case when BKZ reduction is high, and the secret has a lot of non-zero cool bits.

Previous research assumes that once the cruel bits are known, cool bit recovery is easy, and proposes a linear regression recovery method~\citep{benchmarking}. Once the cruel bits are guessed, the quantity $b_{cool}=b - \mathbf{a}_{cruel} \cdot \mathbf s_{cruel}=\mathbf a_{cool} \cdot \mathbf s_{cool} + \epsilon$ is computed ($\mathbf s_{cool/cruel}$ represent the restriction of $\mathbf s$ to the cool/cruel coordinates), and linear regression is used to predict $\mathbf s_{cool}$ from $\mathbf a_{cool}$ and $b_{cool}$.

The use of linear regression, here, is dubious for two reasons. First, we know that the coordinates of $\mathbf a$ are not correlated, and that the secret bits are independent (see Appendix~\ref{app:synthetic}). Linear regression, on the other hand, will compute and invert the test sampled covariance matrix $\mathbf {A}^T\mathbf{A}$, which will have non diagonal elements due to population error that will be amplified by matrix inversion. Second, linear regression ignores the fact that the dot product is computed modulo $q$. As a result, it underestimates the contributions of non-zero bits in the secret (which can ``wrap'' to zero when their sum exceeds $q$). Summarizing, we believe that cool bits should better be recovered one by one rather than all at once, and zero bits are easier to recover that non-zero bits.

For this reason, we propose a new method for cool bit recovery, based on stepwise regression~\citep{efroymson1960multiple}. Once the cruel bits have been guessed, their contribution is subtracted from $b$, and we compute $b_{cool}=b - \mathbf{a}_{cruel} \cdot \mathbf s_{cruel}$. As in the previous method, we run a linear regression on remaining cool bits, but look for the feature with the \emph{lowest contribution}, and assign the value zero to the corresponding secret bit. We normalize coefficients by their maximum absolute value and identify the smallest as zero. We then remove this bit from $(\mathbf a_{cool},b_{cool})$, and repeat the process until we get the known value of $h$ (or a value in a predefined range, if $h$ is not assumed to be known). The key advantage is exploiting sparsity by eliminating zero bits iteratively, avoiding error accumulation from modular arithmetic. 

Stepwise regression recovers the zero bits of the secrets. At first, because the secret is sparse, there are more zeroes than ones, but after a number of steps, the remaining bits of the secrets are mostly ones. In that situation, it is more efficient to consider the dual problem: flip all the remaining cool bits, and perform the regression on $b_{dual} = \mathbf a_{cool} (\mathbf{1}-\mathbf s_{cool}) = \mathbf{a}_{cool}^{\top} \mathbf{1} - b_{cool}$. We call this variant \emph{dual stepwise regression}: we run stepwise regression until the undiscovered cool bits have more ones than zeroes. Then, we alternate between direct and dual recovery.
For instance, if we know, or estimate that, the secret has $10$ cool zeros and $5$ ones, we will apply the direct step $6$ times, before alternating dual and direct. We provide the pseudocode in Appendix~\ref{app:stepwise_reg}.

Tables~\ref{tab:stepfull_256_20} and~\ref{tab:stepfull_512_41} compare linear, stepwise and dual stepwise regression in the two settings with large BKZ reduction (results for the other settings can be  found in Appendix~\ref{app:linear_reg}). We consider secrets with $4$ to $8$ cruel bits, and set the Hamming weight, so that the ratio of cruel bits over $h$ is constant, and equal to $c/n$ for this setting. For each value of $h$ we run models on $20$ different secrets, and report the number of secrets recovered (assuming cruel bits are known), for different numbers of LWE examples used for recovery. Overall, stepwise regression outperforms linear regression, and dual stepwise regression achieves the best results. 

\begin{table}[!h]
    \centering
    \small
    \caption*{\bf Cool bit recovery with Linear, Stepwise, and Dual Stepwise Regression, with 2 different data budgets, out of 20 secrets.}

    \begin{minipage}[t]{0.48\textwidth}
        \centering
        \caption{\bf $\boldsymbol{n=256}, \boldsymbol{\log_2 q=20}$}
        \label{tab:stepfull_256_20}
        \setlength{\tabcolsep}{5pt}
        \begin{tabular}{c|cc|cc|cc}
            \toprule
             Cool bits & \multicolumn{2}{|c}{Linear} & \multicolumn{2}{|c}{Stepwise} & \multicolumn{2}{|c}{Dual} \\
            (Total $h$) & 2M  & 20M   & 2M   & 20M  & 2M  & 20M \\
            \midrule
            26 (30) & 2 & 13 & 3 & 20 & 13 & 20 \\
            32 (37) & 0 & 5 & 0 & 20 & 5 & 20 \\
            38 (44) & 0 & 1 & 0 & 11 & 1 & 20 \\
            45 (52) & 0 & 0 & 0 & 3 & 0 & 18 \\
            52 (60) & 0 & 0 & 0 & 1 & 0 & 14 \\
            \bottomrule
        \end{tabular}
    \end{minipage}
    \hfill
    \begin{minipage}[t]{0.48\textwidth}
        \centering
        \caption{\bf $\boldsymbol{n=512}, \boldsymbol{\log_2 q=41}$}
        \label{tab:stepfull_512_41}
        \setlength{\tabcolsep}{5pt}
        \begin{tabular}{c|cc|cc|cc}
            \toprule
             Cool bits & \multicolumn{2}{|c}{Linear} & \multicolumn{2}{|c}{Stepwise} & \multicolumn{2}{|c}{Dual} \\
            (Total $h$) & 1M  & 4M  & 1M   & 4M   & 1M  & 4M  \\
            \midrule
            40 (44) & 20 & 20 & 20 & 20 & 20 & 20 \\
            50 (55) & 20 & 20 & 20 & 20 & 20 & 20 \\
            60 (66) & 13 & 18 & 17 & 20 & 20 & 20 \\
            70 (77) & 6 & 17 & 15 & 19 & 20 & 20 \\
            80 (88) & 0 & 12 & 10 & 18 & 19 & 20 \\
            \bottomrule
        \end{tabular}
    \end{minipage}
\end{table}

In both settings, dual stepwise regression allows to recover secrets with $8$ cruel bits. For $n=256, \log_2 q=20$,
linear regression cannot recover secrets with more than $4$ cruel bits with $2$ million LWE examples, 
and $6$ with $20$ million. Dual stepwise regression recovers $6$ with $2$ million, and $8$ with $20$ million. For 
$n=512$, $\log_2 q=41$, dual recovery allows for the recovery of $19$ out of $20$ secrets with $8$ cruel bits, with only 
$1$ million LWE examples. This clearly demonstrates the benefits of stepwise regression.

\section{Overall secret recovery}\label{sec:results}

In this section, we present the overall results of our attack, for the four settings. We consider two measures of success. First, we consider the maximum Hamming weight and number of cruel bits that can be recovered, as a function of the training set size and the level of repetition. Then, for a given Hamming weight, we estimate the proportion of all secrets that our method can recover (depending on the number of cruel and cool bits each secret has). In these experiments, we assume that the attacker only knows the Hamming weight $h$ of the secret to be attacked. In particular, the attacker does not know the cruel bits. We generate 15,000 model-based attempts of the cruel bits, and use dual stepwise recovery for cool bit recovery. 

For $n=256$, $\log_2 q=12$, 
SALSA recovers secrets with $h=8$, and Cool\&Cruel with $h=12$. We recover secrets with $h=14$, and $5$ cruel bits (Table ~\ref{tab:results_256_12_a}). Our best results are achieved with small data budgets ($4$ million examples) and large repetition ($15$ times). Note that models trained on BKZ-reduced data achieve the same results as those trained on synthetic data. 
For $n=512$, $\log_2 q=28$ (Table ~\ref{tab:results_512_28_a}), the other hard setting (small modulus, larger cruel region), we recover $h=12$ (and $5$ cruel bits), matching Cool\&Cruel. Again, performance on reduced and synthetic data is the same, and repetition matters more than the number of reduced examples.

\begin{table}[h]
\centering
\small
\caption*{\bf Highest Hamming weight/cruel bits $h/k$ recovered: binary secret, harder setting.}

    \begin{minipage}[t]{0.48\textwidth}
        \centering
        \caption{\bf $\boldsymbol{n=256}, \boldsymbol{\log_2 q=12}$}
        \label{tab:results_256_12_a}
        \begin{tabular}{l|ccccc}
            \toprule
             & \multicolumn{5}{|c}{Repetition} \\
               & 1x & 2x & 5x & 15x & 50x \\
            \midrule
            \multicolumn{3}{l}{BKZ-reduced data} \\
            4M  & 10/3 & 10/3 & 12/4 & \textbf{14/5} & 14/5 \\
            \midrule
            \multicolumn{3}{l}{Synthetic data} \\
            4M  & 10/3 & 10/3 & 10/3 & \textbf{14/5} & 14/5 \\
            20M  & 10/3 & 10/3 & 10/3 & 14/4 & 14/5 \\
            50M  & 10/3 & 10/3 & 12/3 & 14/5 & - \\
            100M & 10/4 & 10/4 & 12/5 & - & - \\
            200M & 12/5 & 12/5 & 12/5 & - & - \\
            400M & 12/5 & 12/5 & - & - & - \\
           \bottomrule
            \multicolumn{6}{c}{Best of 80 models.} \\
        \end{tabular}
    \end{minipage}
    \hfill
    \begin{minipage}[t]{0.48\textwidth}
        \centering
        \caption{\bf $\boldsymbol{n=512}, \boldsymbol{\log_2 q=28}$}
        \label{tab:results_512_28_a}
        \begin{tabular}{l|ccccc}
            \toprule
             & \multicolumn{5}{|c}{Repetition} \\
             & 1x & 2x & 5x & 15x & 50x \\
            \midrule
            \multicolumn{3}{l}{BKZ-reduced data} \\
            4M  & 10/3 & 10/3 & 10/4 & 12/4 & 12/4 \\
            20M  & 10/4 & 10/4 & 10/5 & \textbf{12/5} & 12/5 \\
            50M  & 10/4 & 10/4 & 10/3 & 12/5 & - \\
            \midrule
            \multicolumn{3}{l}{Synthetic data} \\
            4M  & 10/3 & 10/3 & 10/3 & 10/3 & 12/4 \\
            20M  & 10/3 & 10/4 & \textbf{12/5} & 12/5 & 12/5 \\
            50M  & 10/3 & 10/4 & 12/4 & 12/4 & - \\
            100M & 10/3 & 12/4 & 12/5 & - & - \\
            200M & 12/4 & 12/4 & 12/4 & - & - \\
           \bottomrule
            \multicolumn{6}{c}{Best of 80 models.} \\
        \end{tabular}
    \end{minipage}
\end{table}

\begin{table}[h]
\centering
\small
\caption*{\bf Highest Hamming weight/cruel bits $h/k$ recovered: binary secret, easier setting.}

    \begin{minipage}[t]{0.48\textwidth}
        \centering
        \caption{\bf $\boldsymbol{n=256}, \boldsymbol{\log_2 q=20}$}
        \label{tab:results_256_20_a}
        \begin{tabular}{l|ccccc}
            \toprule
             & \multicolumn{5}{|c}{Repetition} \\
            & 1x & 2x & 5x & 15x & 50x \\
            \midrule
            \multicolumn{3}{l}{BKZ-reduced data} \\
            4M  & 55/6 & 55/6 & 55/7 & 60/8 & 65/8 \\
            20M & 55/6 & 55/6 & 60/7 & 60/8 & 65/8 \\
            50M  & 60/8 & 65/8 & 65/8 & 65/8 & - \\
            100M  & 65/8 & 65/8 & \textbf{70/8} & - & - \\
            200M  & 65/8 & 70/8 & 70/8 & - & - \\
            400M  & 65/8 & 70/8 & - & - & - \\
            \midrule
            \multicolumn{3}{l}{Synthetic data} \\
            4M  & 55/6 & 60/7 & 60/7 & 60/7 & 60/8 \\
            20M  & 60/8 & 65/8 & 65/8 & 65/8 & 65/8 \\
            50M  & 65/7 & 65/8 & \textbf{70/8} & 70/8 & - \\
            100M  & 65/8 & 65/8 & 70/8 & - & - \\
            200M  & 65/8 & 65/8 & 70/8 & - & - \\
            400M  & 65/8 & 65/8 & - & - & - \\
            \bottomrule
            \multicolumn{6}{c}{Best of 80 models.} \\
        \end{tabular}
    \end{minipage}
    \hfill
    \begin{minipage}[t]{0.48\textwidth}
        \centering
        \caption{\bf $\boldsymbol{n=512}, \boldsymbol{\log_2 q=41}$}
        \label{tab:results_512_41_a}
        \begin{tabular}{l|cccccc}
            \toprule
             & \multicolumn{5}{|c}{Repetition} \\
             & 1x & 2x & 5x & 15x & 50x \\
            \midrule
            \multicolumn{3}{l}{BKZ-reduced data} \\
            4M  & 70/4 & 70/4 & 70/4 & 70/6 & 70/6 \\
            \midrule
            \multicolumn{3}{l}{Synthetic data} \\
            4M  & 70/4 & 70/4 & 70/4 & 70/5 & 70/6 \\
            20M & 70/6 & 70/6 & 70/7 & \textbf{75/7} & 75/7 \\
            50M  & 70/6 & 70/6 & 70/6 & 75/7 & - \\
            100M  & 70/6 & 70/6 & 70/6 & - & - \\
            200M  & 70/6 & 70/6 & - & - & - \\
            \bottomrule
            \multicolumn{6}{c}{Best of 80 models.} \\
        \end{tabular}
    \end{minipage}
\end{table}

The improvement over prior work is larger in settings with more reduction, i.e. smaller cruel regions. For $n=256$, $\log_2 q=20$, we recover secrets with $h=70$ and $8$ cruel bits, vs $h=33$ for VERDE (Table ~\ref{tab:results_256_20_a}). For $n=512$, $\log_2 q=41$, we recover secrets with $h=75$ and $7$ cruel bits, vs $63$ in VERDE, and $60$ in Cool\&Cruel (Table ~\ref{tab:results_512_41_a}). As before, there is little difference between models trained on reduced and synthetic data. In Appendix ~\ref{app:ternary} we show \textbf{similar results on ternary secrets}: for $n=256$, $\log_2 q=20$, we recover secrets with $h=55$ and $8$ cruel bits, compared to $h=24$ from VERDE. Similarly, for $n=512$, $\log_2 q=41$, we recover secrets with $h=75$ and $7$ cruel bits vs $66$ in FRESCA.

The metric used so far, the largest recoverable Hamming weight, can be misleading. A secret with a very large Hamming weight could, with a very low probability, have a very small number of cruel bits, and be recoverable. We therefore consider the proportion of all secrets with a given $h$ that our attack can recover. 

For any value of $h$ and $k$, the number of cruel bits in the secret, we compute our model recovery rate $r(h,k)$, and the probability $p(h,k)$ that a random secret has $k$ cruel bits, which follows a hyper-geometric distribution. Then, the expected recovery rate is $\mathcal E(h)=\sum_{k=0}^h p(h,k)r(h,k).$

\begin{table}[!h]
\centering
\small
\caption*{\bf Expected recovery rate for different Hamming weights.}

    \begin{minipage}[t]{0.48\textwidth}
        \centering
        \caption{\bf $\mathcal E(h)$ for $\boldsymbol{n=256}, \boldsymbol{\log_2 q=20}$}
        \label{tab:results_256_20_b}
        \begin{tabular}{c|ccccc}
            \toprule
                $h$ & \multicolumn{1}{c}{33} & \multicolumn{1}{c}{55} & \multicolumn{1}{c}{60} & \multicolumn{1}{c}{65} & \multicolumn{1}{c}{70} \\ \midrule
                \multicolumn{1}{c|}{VERDE} & 33\% & 4\% & 2\% & 1\% & 0\% \\
                \multicolumn{1}{c|}{Ours} & \textbf{98\%} & \textbf{71\%} & \textbf{60\%} & \textbf{49\%} & \textbf{38\%} \\
             \bottomrule
        \end{tabular}
    \end{minipage}
    \hfill
    \begin{minipage}[t]{0.48\textwidth}
        \centering
        \caption{\bf $\mathcal E(h)$ for $\boldsymbol{n=512}, \boldsymbol{\log_2 q=41}$}
        \label{tab:results_512_41_b}
        \begin{tabular}{c|cccc}
            \toprule
                $h$ & \multicolumn{1}{c}{63} & \multicolumn{1}{c}{65} & \multicolumn{1}{c}{70} & \multicolumn{1}{c}{75} \\ \midrule
                \multicolumn{1}{c|}{VERDE} & 15\% & 14\% & 10\% & 7\% \\
                \multicolumn{1}{c|}{Ours} & \textbf{91\%} & \textbf{89\%} & \textbf{72\%} & \textbf{63\%} \\
             \bottomrule
        \end{tabular}
    \end{minipage}
\end{table}

In Tables~\ref{tab:results_256_20_b} and~\ref{tab:results_512_41_b}, we compare expected recovery rate $\mathcal E(h)$ for our method and VERDE. Our method not only recovers higher $h$, but it is much more reliable. For $n=256$, $\log_2 q=20$ and $h=33$, VERDE recovers $33\%$ of secrets (up to $3$ cruel bits), while we recover \textbf{98\%} (up to $8$ cruel bits). For $n=512$, $\log_2 q=41$ and $h=63$, VERDE recovers $15\%$ of secrets, while we recover \textbf{91\%}.

\section{Scaling laws}

We define model-based attempts $A$ as the number of attempts needed to get the correct cruel bits (attempts are generated from the distinguisher output). We choose $A$ as a more fine-grained metric of model performance on secret recovery. For $n=256$, $\log_2 q=20$, in the binary case, we present empirical laws relating the number of model-based attempts, $A$, required to recover a secret for model size $N$, data amount $D$, and repetitions $R$ for a fixed secret $s$ with Hamming weight $h$. 

\noindent \textbf{Model parameters law:} To understand any scaling pattern between the model size $N$ and model-based attempts $A$, we vary the embedding dimension between $256$ and $1024$ and the number of layers between $4$ and $12$. We test different model sizes for three different secrets with Hamming weights $h=\{60, 65, 70\}$. We use $100$ million data and 1 repetition. As shown in Figure ~\ref{fig:model}, the number of model-based attempts is not improved by an increase in model parameters. We leave the exploration of larger models for future work.

\begin{figure}[h]
    \centering

    \begin{minipage}[t]{0.48\textwidth}
        \centering
        \begin{tikzpicture}
        \begin{loglogaxis}[
            width=\textwidth, height=5.5cm,
            xlabel={Model params $N$},
            ylabel={Model based attempts $A$},
            legend style={font=\small, at={(0.05,0.42)}, anchor=south west, draw=none, fill opacity=0.6, text opacity=1}
        ]
        \addplot+[only marks, mark=triangle*, thick, color=C0] table {
        N G
        3145728 2415
        4718592 3517
        6291456 2721
        9437184 4268
        7077888 2983
        10616832 15712
        14155776 3069
        21233664 4597
        12582912 3566
        18874368 4675
        25165824 6439
        37748736 5305
        19660800 5149
        29491200 1681
        39321600 2038
        58982400 5048
        28311552 7851
        42467328 7629
        56623104 6753
        84934656 2549
        38535168 5585
        57802752 4803
        77070336 3641
        115605504 6253
        50331648 5054
        75497472 3849
        100663296 1406
        150994944 1908
        };
        \addlegendentry{$h=60$}
        \addplot+[only marks, mark=triangle*, thick, color=C1] table {
        N G
        3145728 13888
        4718592 14382
        6291456 15495
        9437184 15216
        7077888 9223
        10616832 20806
        14155776 23527
        21233664 37003
        12582912 14073
        18874368 14210
        25165824 36197
        37748736 25499
        19660800 7984
        29491200 14543
        39321600 14732
        58982400 12288
        28311552 19038
        42467328 17014
        56623104 22604
        84934656 13815
        38535168 24577
        57802752 18841
        77070336 16707
        115605504 14689
        50331648 12053
        75497472 13239
        100663296 6560
        150994944 14454
        };
        \addlegendentry{$h=65$}
        \addplot+[only marks, mark=triangle*, thick, color=C2] table {
        N G
        3145728 1519158
        4718592 1560657
        6291456 1073151
        9437184 1576291
        7077888 1704112
        10616832 1820944
        14155776 501867
        21233664 1781547
        12582912 800021
        18874368 653333
        25165824 426274
        37748736 791729
        19660800 733426
        29491200 587639
        39321600 757877
        58982400 1104624
        28311552 1900899
        42467328 1103631
        56623104 2796479
        84934656 2617011
        38535168 2577302
        57802752 899561
        77070336 1606322
        115605504 1456441
        50331648 1098916
        75497472 1563086
        100663296 1178450
        150994944 883732
        };
        \addlegendentry{$h=70$}
        \addplot+[domain=3e6:2e8, samples=2, thick, color=C0, opacity=0.8, no markers, const plot, smooth] { (6012) / x^(0.0229) };
        \addplot+[domain=3e6:2e8, samples=2, thick, color=C1, opacity=0.8, no markers, const plot, smooth] { (2.252e4) / x^(0.0199) };
        \addplot+[domain=3e6:2e8, samples=2, thick, color=C2, opacity=0.8, no markers, const plot, smooth] { (7.829e5) / x^(-0.0240) };
        \end{loglogaxis}
        \end{tikzpicture}
        \caption{\textbf{Model parameters $N$ vs model based attempts $A$ for three secrets with different Hamming weights $h$.}}
        \label{fig:model}
    \end{minipage}
    \hfill
    \begin{minipage}[t]{0.48\textwidth}
        \centering
        \begin{tikzpicture}
        \begin{loglogaxis}[
            width=\textwidth, height=5.5cm,
            xlabel={Total training data $D$},
            ylabel={Model based attempts $A$},
            legend style={font=\small, at={(0.05,0.05)}, anchor=south west, draw=none, fill opacity=0.6, text opacity=1}
        ]
        \addplot+[only marks, mark=triangle*, thick, color=C0] table {
        N G
        1000000 63594037
        1250000 50226830
        1500000 36351155
        1750000 41126407
        2000000 27842180
        2500000 17241888
        3000000 9842651
        3500000 6819337
        4000000 7682518
        5000000 5638639
        6000000 7707003
        7000000 8296726
        8000000 12112258
        10000000 6871977
        12500000 2038247
        15000000 4632962
        17500000 2164880
        20000000 1587806
        25000000 1037272
        30000000 3134557
        35000000 1698253
        40000000 1196122
        50000000 1063186
        60000000 1306668
        70000000 2018405
        80000000 974811
        100000000 1519158
        125000000 973826
        150000000 1561519
        175000000 1510559
        200000000 912402
        250000000 751996
        300000000 1011239
        350000000 709434
        400000000 494317
        };
        \addlegendentry{$R=1$}

        \addplot+[only marks, mark=triangle*, thick, color=C1] table {
        N G
        2000000 13011025
        2500000 10436317
        3000000 19902238
        3500000 10130343
        4000000 5087317
        5000000 5890647
        6000000 3450777
        7000000 2747444
        8000000 4194957
        10000000 3103636
        12000000 1573348
        14000000 1308914
        16000000 695949
        20000000 1046229
        25000000 565507
        30000000 706015
        35000000 369674
        40000000 241750
        50000000 127714
        60000000 132104
        70000000 95362
        80000000 62016
        100000000 43602
        120000000 64822
        140000000 39942
        160000000 28830
        200000000 41139
        250000000 24767
        300000000 19022
        350000000 27105
        400000000 12455
        500000000 20871
        600000000 13307
        700000000 11350
        800000000 11060
        };
        \addlegendentry{$R=2$}

        \addplot+[only marks, mark=triangle*, thick, color=C2] table {
        N G
        5000000 6805233
        6250000 5835247
        7500000 3713013
        8750000 3696540
        10000000 3359180
        12500000 2634705
        15000000 3034115
        17500000 1226211
        20000000 553301
        25000000 501521
        30000000 396357
        35000000 521483
        40000000 156242
        50000000 118560
        62500000 101004
        75000000 111916
        87500000 78983
        100000000 110024
        125000000 79144
        150000000 57569
        175000000 42859
        200000000 33953
        250000000 29795
        300000000 19701
        350000000 19398
        400000000 11484
        500000000 6986
        625000000 4744
        750000000 3460
        875000000 4862
        1000000000 5237
        };
        \addlegendentry{$R=5$}

        \addplot+[only marks, mark=triangle*, thick, color=C3] table {
        N G
        15000000 19290234
        18750000 15048769
        22500000 11741615
        26250000 11872291
        30000000 6701820
        37500000 5314214
        45000000 1888316
        52500000 1863797
        60000000 745748
        75000000 703873
        90000000 523177
        105000000 594018
        120000000 203503
        150000000 405694
        187500000 363936
        225000000 309051
        262500000 433474
        300000000 180107
        375000000 175799
        450000000 68293
        525000000 40083
        600000000 46579
        750000000 35607
        };
        \addlegendentry{$R=15$}

        \addplot+[only marks, mark=triangle*, thick, color=C4] table {
        N G
        50000000 24248034
        62500000 14590417
        75000000 8733864
        87500000 9970966
        100000000 3933386
        125000000 2728981
        150000000 2105741
        175000000 1439184
        200000000 972952
        250000000 498651
        300000000 217399
        350000000 306950
        400000000 174390
        500000000 330853
        625000000 101087
        750000000 98150
        875000000 116628
        1000000000 92014
        };
        \addlegendentry{$R=50$}

        \addplot+[domain=1e6:1e9, samples=2, thick, color=C0, opacity=0.8, no markers, const plot, smooth] { (4.92e11) / x^(0.7031) };
        \addplot+[domain=2e6:1e9, samples=2, thick, color=C1, opacity=0.8, no markers, const plot, smooth] { (3.03e15) / x^(1.3142) };
        \addplot+[domain=5e6:1e9, samples=2, thick, color=C2, opacity=0.8, no markers, const plot, smooth] { (3.701e16) / x^(1.4529) };
        \addplot+[domain=15e6:1e9, samples=2, thick, color=C3, opacity=0.8, no markers, const plot, smooth] { (3.766e18) / x^(1.5844) };
        \addplot+[domain=50e6:1e9, samples=2, thick, color=C4, opacity=0.8, no markers, const plot, smooth] { (1.849e22) / x^(1.9507) };

        \end{loglogaxis}
        \end{tikzpicture}
        \caption{\textbf{Total training data $D$ and repetition $R$ vs model based attempts $A$ for one secret with Hamming weight $h=70$. Total distinct data can be computed as $D/R$.}}
        \label{fig:data}
    \end{minipage}
\end{figure}

\noindent \textbf{Data-repetition law:} We fix the model embedding dimension at 256 and 4 layers. We vary the amount of distinct data from 1 to 400 million, and data repetition $R$ from 1 to 50. We define $D$ as the total training data, equal to distinct training data times the data repetition factor $R$. Figure ~\ref{fig:data} presents the results for the best-performing model out of 8 different initializations for one secret with $h=70$; broader validation is left for future work. We propose the following functional form: $\ln(A_R) = C_R - \alpha_R \ln(D)$ and we report the fitted parameters using least square errors between $(\ln(D), \ln(A_R)$) across 5 distinct $R$ regimes in Table ~\ref{tab:scaling_law_params}. Notably, our experiments reveal that $\alpha_1$ is considerably lower than $\alpha_R$, $R>1$. {Therefore, data repetition does not just diminish the required number of distinct (costly) samples but actually reshapes the scaling power law of model based attempts $A$ as a function of $D$.} Repeated data is crucial for recovering a secret with Hamming weight $h=70$. \textbf{Simply increasing data is important but insufficient}; a careful data strategy is necessary to tackle the LWE problem.

\begin{table}[h]
    \small
    \centering
    \setlength{\tabcolsep}{4pt}
    \begin{tabular}{c|ccccc}
        \toprule
         $R$ & 1 & 2 & 5 & 15 & 50 \\
        \midrule
         $C_R$ & 26.9 & 35.6 & 38.1 & 42.8 & 51.3 \\
         $\alpha_R$ & 0.70 & 1.31 & 1.45 & 1.58 & 1.95\\
         CI & 0.61 - 0.79 & 1.25 - 1.4 & 1.38 - 1.51 & 1.44 - 1.71 & 1.76 - 2.26 \\
       \bottomrule
    \end{tabular}
    \caption{Empirical scaling power law fitted parameters. CI (95\% Confidence Interval for $\alpha_R$) is computed using bootstrapping.}
    \label{tab:scaling_law_params}
\end{table}

\section{Discussion and conclusion}

We introduced three main techniques for improving secret recovery in ML attacks: using larger training sets, repeating training examples, and stepwise regression for cool bits recovery. This brings considerable improvement over previous attacks, both in terms of the maximum recoverable Hamming weight and the proportion of secrets recovered for a given $h$. We additionally presented an empirical scaling power law that relates model based attempts to data amount and data repetitions.

{Limitations.} Our attack targets Learning With Errors in its basic form. We make no claim about its applicability to other PQC algorithms, but believe it can be adapted to the variants of LWE considered in the SALSA papers (Ring and Module LWE, notably). Our experiments focus on sparse binary and ternary secrets, but the attack can be adjusted to small secrets, following the methods described in VERDE~\citep{verde}. Finally, we target sparse secrets: with about $5\%$ density for the harder settings, and $14$ and $27\%$ in the easier ones.

Our conclusions extend in two directions. First, larger training sets, used for many epochs, are key to recovering the cruel bits. This confirms previous observations about the importance of repeated data~\citep{charton2024emergentpropertiesrepeatedexamples,saxena2025making}. The need for large training sets, and data for stepwise regression, may limit the applicability of our attack: whereas large reduced datasets can be created from small eavesdropped samples~\citep{picante}, BKZ reduction is costly, both in terms of time and computing resources. We believe synthetic data can be used to simulate experimental settings, without the costly preprocessing step. This has two key implications: (1) it allows synthetic data to be used for experimentally establishing scaling laws for breaking arbitrary binary, ternary, or small secrets and (2) it suggests that ML attacks could be improved using pretraining on synthetic data, which is essentially free to generate compared to data reduction via lattice reduction techniques.

Our second conclusion is the benefit of stepwise regression. Most statistical handbooks consider stepwise regression a subpar alternative~\citep{stayaway}, because it does not take into account possible correlations between input features. We believe our conclusions stem from the very specific nature of the data considered here: the columns of matrix $\mathbf A$ are uncorrelated, even after BKZ-reduction, and stepwise regression enforces this inductive bias. In most problems of statistics, on the other hand, features can be correlated, and stepwise regression is harmful. 

Our results improve the performance of ML-based attacks on LWE and outperform non-ML strategies at comparable budgets and Hamming weights. {By investigating scaling laws on LWE for the first time, we provide some insight for which strategies can be used to tackle harder versions of the LWE problem.} On a broader level, this line of research is important, because PQC systems like LWE are the future standard for safe digital transactions, and the community must have a clear understanding of their potential limitations and weaknesses, notably those relative to the use of sparse and small secrets, before they are deployed at a large scale. Weaknesses discovered now, and taken into account in the standards, may prevent future attacks against PQC.

\newpage

\section*{Acknowledgements}
The authors would like to thank Mohamed Malhou for his valuable input and feedback on this work.

\bibliography{bib}
\bibliographystyle{bib}

\newpage

\appendix
\onecolumn
\section*{Appendix}

\section{BKZ-reduced dataset}\label{app:bkz}

We report some statistics in Table~\ref{tab:red_params} of the BKZ-reduced datasets. $c$ is the size of the cruel region (coordinates with variance larger than half the variance of the uniform law), $\sigma_{cool}$ is the standard deviation of the coordinates of the cool region centered in $[-q/2, q/2]$, $\rho$ is the mean of absolute values of off-diagonal pairwise correlation from the covariance matrix of the reduced samples, and $\sigma_\epsilon$ is the standard deviation of $\mathbf R \epsilon$ centered in $[-q/2, q/2]$.

It is important to note that reduction creates a cool region of size $n-c$ but introduces an error term with standard deviation approximately equal to $q/\sqrt{12}$ (the standard deviation of one cruel bit). 

\begin{table}[h]
\caption{{\bf Reduction statistics}. We report dataset size, CPU hours, $c$ (size of the cruel region), $\sigma_{cool}$ (standard deviation of the cool region), $\rho$ (average non-diagonal pairwise absolute value correlation from the covariance matrix of $\mathbf {RA}$), $\sigma_\epsilon$ (standard deviation of $\mathbf R \epsilon$).}
\label{tab:red_params}
\small
\centering
\begin{tabular}{cccccccc}
\toprule
$n$ & $\log_2q$ & Dataset size & CPU hours & $c$ & $\sigma_{cool}$ & $\rho$ & $\sigma_e$ \\
& & (millions) & (millions) & & &  \\
\midrule
256 & 12 & 4 & 0.1 & 143 & $0.30q/\sqrt{12}$ & $0.18\%$ & $0.88q/\sqrt{12}$ \\
256 & 20 & 400 & 40.4 & 34 & $0.23q/\sqrt{12}$ & $1.05\%$ & $0.90q/\sqrt{12}$ \\
512 & 28 & 60 & 1.0 & 224 & $0.19q/\sqrt{12}$ & $0.12\%$ & $0.70q/\sqrt{12}$ \\
512 & 41 & 4 & 0.1 & 46 & $0.15q/\sqrt{12}$ & $0.18\%$ & $0.80q/\sqrt{12}$ \\
\bottomrule    
\end{tabular}%
\end{table}

\section{Synthetic dataset}\label{app:synthetic}

Some of the experiments in the manuscript require very large training sets (up to $400$ million reduced data), which requires a lot of computing resources. We reduce $400$ million LWE samples for $n=256$ $\log_2 q=20$, but for the three other settings, we relied on synthetic data to complement smaller BKZ-reduced datasets (see Table~\ref{tab:red_params}). 

To generate synthetic data, we observe that after reduction the coordinates are uncorrelated (i.e. the off-diagonal terms of the covariance matrix $(\mathbf{RA})^T(\mathbf{RA})$ are very small), i.e. the mean absolute value $\rho$ is close to 0. This suggests a method for generating synthetic data. One million LWE samples are reduced using BKZ, and we use these reduced examples to measure $c$, $\sigma_{cool}$ and $\sigma_\epsilon$. The synthetic reduced $\mathbf a$ are then generated by sampling the first $c$ coordinates from a uniform distribution, and the $n-c$ others from a centered discrete Gaussian distribution with variance $\sigma_{cool}$. We then compute $b=\mathbf a \cdot \mathbf s + \epsilon$, with $\epsilon$ a centered discrete Gaussian variable, with standard deviation $\sigma_\epsilon$. Note that because all unreduced columns are assumed to have the variance of the uniform distribution, the synthetic data is a little less reduced than the BKZ-reduced data. 

When BKZ-reduction is applied, the original matrix $\mathbf A$ and vector $\mathbf b$ of LWE samples are subjected to a linear transformation $\mathbf R$. The columns of $\mathbf{RA}$ are not correlated, and have a block structure: the $c$ first column have high variance, and the $n-c$ last columns low variance. Multiplying $\mathbf {RA}$ by any block diagonal matrix of the form 

\begin{equation}
\mathbf{O} =
 \begin{bmatrix}
\Omega_1 & 0 \\
0 & \Omega_2 
\end{bmatrix}
\label{eq:qary1}
\end{equation}

with $\Omega_1$ and $\Omega_2$ two $c\times c$ and $(n-c)\times (n-c)$ matrices, near-orthogonal, (i.e. with condition numbers close to $1$), should have little impact on the distribution of cool and cruel bits. This means we can generate, from BKZ-reduced samples $\mathbf{RA}$, a lot of synthetic samples, with the same secret, reduction factor, and noise, by sampling quasi-orthogonal matrices $\mathbf O$, and generating new data $\mathbf {ORA}$ and $\mathbf{ORb}$. We leave an analysis of this data augmentation method for future work.

\section{Cool and cruel embedding ablation} \label{app:cc_emb_ablation}

For $n=256$ $\log_2 q=20$ we train 16 different models with 400 million samples and 2 repetitions for 5 different secrets with 4 cruel bits. In Table~\ref{tab:cc_emb_table} we compare the recovery rate (i.e. if the secret is fully recovered) with or without the cool and cruel embedding introduced in Section~\ref{sec:settings}. We highlight that the new embedding allows for recoveries on all secrets.

\begin{table}[h]
\caption{{\bf Cool and cruel embedding comparison}.}
\label{tab:cc_emb_table}
\small
\centering
\begin{tabular}{ccc}
\toprule
$h$ & CC embedding disabled & CC embedding enabled \\
\midrule
28 & \textbf{5/16} & \textbf{8/16} \\
30 & 0/16 & \textbf{4/16} \\
33 & \textbf{3/16} & \textbf{7/16} \\
34 & 0/16 & \textbf{5/16} \\
36 & 0/16 & \textbf{1/16} \\
\bottomrule    
\end{tabular}%
\end{table}

\section{Bias in angular embeddings and ways to circumvent it}\label{app:collapse}

To predict $b=\mathbf a\cdot \mathbf s+\epsilon$ from $\mathbf a$, our transformer uses an angular embedding. The model outputs a point $P(x,y)$ on the real plane, the integers $ i \in \{0,\dots,q-1\}$ are encoded as the points $B_i(\cos(\frac {2\pi i} q), \sin(\frac {2\pi i} q))$ on the unit circle, and the model prediction is the point $B_i$ closest to $P$. If $P$ has polar coordinates $(r\cos(\theta), r\sin(\theta))$, with $r>0$, the point $B_{i_0}$ closest to $P$ verifies $i_0 = \mathrm{argmin}_i | \theta - \frac{2\pi i} q|$.

During training, the model learns to minimize a Mean Square Error (MSE) loss. If the angular embedding of $B$ is $(x_0,y_0)$, and if the model predicts $P(x,y)$, the loss is $l=(x-x_0)^2+(y-y_0)^2$. Since the possible values of $b$ are uniformly distributed on the unit circle, we can assume, without loss of generality, that $B=(1,0)$. Therefore, the loss is $$
l=(1-r\cos(\theta))^2+r^2\sin^2(\theta)= 1+r^2-2r\cos(\theta).$$

During the early stages of training, the model has not learned modular arithmetic and $b$ is predicted at random. Suppose that all model predictions lie at a distance $r$ of the origin, the average MSE loss is the integral of $l$ over all possible angles, so $$\mathcal{L}(r)=\frac{1}{2\pi r}\int_{-\pi}^{\pi} (1+r^2-2r\cos(\theta)) r \, d\theta = 1 + r^2.$$

Therefore, for a clueless model (a model that predicts $b$ at random), the average loss is $1+r^2$, and the optimizer can reduce the loss just by making $r$ smaller. Model predictions will therefore tend to collapse towards the origin, at which point the loss becomes constant ($l(0)=1$ no matter the prediction), and nothing can be learned anymore.

Note that collapse only happens when the model cannot predict $b$. If the model learns to predict $b$ up to some error, i.e. that, assuming $B=(1,0)$ the predicted value of $\theta$ lies in the interval $[-\varepsilon,\varepsilon]$ for some small $\varepsilon$, the average loss then becomes: 
$$\mathcal L_{\varepsilon}(r) = 1+r^2-2r\frac{\sin(\varepsilon)}{\varepsilon} = (r-1)^2 + 2r\left(1-\frac{\sin(\varepsilon)}{\varepsilon}\right).$$
This shows that once the model starts learning, model predictions stop collapsing to the origin. In other words, model collapse only happens at the beginning of training. (Note, we assume here that model predictions are uniformly distributed over $[-\varepsilon,\varepsilon]$, this is a simplification. It can be shown that the same phenomenon appears if the distribution of predictions is unimodal, and centered on $0$).

To prevent the model from collapsing in the initial phase of training, we add to the loss a penalty $\mathcal P(r) = \alpha r^2 +\frac \beta {r^2}$. The average loss (for a random prediction of $b$) then becomes $$\mathcal L(r) = 1+(1+\alpha) r^2 + \frac \beta {r^2}.$$
It reaches a minimum for $\mathcal L'(r) = 2(1+\alpha)r-2\beta / r^3=0$ or $r^4={\frac \beta{1+\alpha}}.$

In Table~\ref{tab:loss_table} we experiment with two different settings and we report the hardest recovered secret by the two different settings. The first setting fixes $\beta=1$ and $\alpha=0$ to force predictions to remain close to the unit circle in the initial, ``clueless'', phase of learning. The second setting is inherited from Saxena et al.~\citep{saxena2025making}, where authors suggest $\alpha=\beta=0.1$. 

Overall the two settings are similar, with the second setting being more successful or at par on almost all data budgets.

\begin{table}[h]
\caption{{\bf Loss comparison}. }
\label{tab:loss_table}
\small
\centering
\begin{tabular}{cccc}
\toprule
Data budget & Repetitions & $\alpha=0, \beta=1$ & $\alpha=0.1, \beta=0.1$ \\
\midrule
\multicolumn{4}{l}{$N=256$ $\log_2 q = 20$} \\
50M & 15x & 60/8 & 65/8 \\
100M & 5x & 65/8 & 65/8 \\
200M & 5x & 65/8 & 65/8 \\ 
400M & 2x & 65/8 & \textbf{70/8} \\
\midrule
\multicolumn{4}{l}{$N=512$ $\log_2 q = 28$} \\
20M & 15x & 12/4 & \textbf{12/5} \\
50M & 15x & 12/4 & 12/4 \\
\bottomrule    
\end{tabular}%
\end{table}

\section{Stepwise regression algorithm}\label{app:stepwise_reg}

We include the pseudocode to replicate the stepwise regression algorithm. Note: $j^*$ is a local index; mapping to global index uses the \texttt{active} vector. Variable $b_{primal}$ maintains the primal residual for mode switching.

\begin{algorithm}[!ht]
\DontPrintSemicolon
\KwIn{Matrix $\mathbf a_{cool} \in \mathbb{Z}_q^{m \times cool}$, vector $b_{cool} \in \mathbb{Z}_q^{m}$, int $h_{cool}$, flag \texttt{use\_dual\_algo}}
\KwOut{Secret guess $g$}

$g \gets$ vector of length $cool$ with all entries equal to $0$\;
$active \gets$ vector of length $cool$ with all entries equal to $1$\;
$b_{primal} \gets b_{cool}$\;
$ones \gets h_{cool}$\;
$zeros \gets cool - h_{cool}$\;

\While{(\texttt{use\_dual\_algo} $\wedge$ $|active| > 0$) \textbf{or} ($\neg$\texttt{use\_dual\_algo} $\wedge$ $|active| > h_{cool}$)}{
    $use\_dual \gets (ones > zeros) \wedge \texttt{use\_dual\_algo}$\;

    \tcp{One-step regression}
    $b \gets use\_dual$ ? $(\mathbf a_{cool}^T\mathbf{1} - b_{primal}) \bmod q$ : $b_{primal}$\;

    $X \gets \mathbf a_{cool}[:, active]$\;
    $y \gets b$\;
    $coef \gets \arg\min_c \|y - Xc\|_2^2$\;
    $coef \gets coef / \max(|coef|)$\;
    
    \tcp{Remove weakest feature}
    $j^* \gets \arg\min_j |coef_j|$\;
    $j^*_{g} \gets$ global index of $j^*$-th active column\;
    $active[j^*_{g}] \gets 0$\;

    \If{$use\_dual$}{
        $b_{primal} \gets (b_{primal} - \mathbf a_{cool}[:, j^*_{g}]) \bmod q$\;
        $g[j^*_{g}] \gets 1$\;
        $ones \gets ones - 1$\;        
    }
    \Else{
        $zeros \gets zeros - 1$\;
    }
}
\If{$\neg$\texttt{use\_dual\_algo}}{
    $g[active] \gets 1$
}
\Return{$g$}
\caption{Linear Secret Backward Reduction}
\end{algorithm}

\section{Linear regression}\label{app:linear_reg}

Similar to Section~\ref{sec:stepwise} we compare linear, stepwise and dual stepwise regression and we report the cool bits recovery for the two harder settings, where the BKZ-reduced data has a larger cruel region. As shown in Tables~\ref{tab:step_256_12}, ~\ref{tab:step_256_20}, ~\ref{tab:step_512_28} and ~\ref{tab:step_512_41} dual stepwise regression shows the best performance. 

\begin{table}[!h]
    \small
    \centering
    \caption{{\bf Cool bits recovery $\boldsymbol{n=256}, \boldsymbol{\log_2 q=12}$} assuming cruel bits have been recovered}
    \begin{tabular}{cc|ccc|ccc|ccc}
        \toprule
         & & \multicolumn{3}{|c}{Linear regression} & \multicolumn{3}{|c}{Stepwise regression} & \multicolumn{3}{|c}{Dual stepwise regression} \\
        Cool bits & Total $h$ & 1M   & 2M  & 4M  & 1M   & 2M   & 4M   & 1M  & 2M  & 4M  \\
        \midrule
        8 & 18 & 20 & 20 & 20 & 20 & 20 & 20 & 20 & 20 & 20 \\
        18 & 41 & 8 & 20 & 20 & 20 & 20 & 20 & 20 & 20 & 20 \\
        23 & 52 & 0 & 3 & 7 & 0 & 9 & 20 & 13 & 20 & 20 \\
        28 & 63 & 0 & 0 & 0 & 0 & 0 & 2 & 0 & 3 & 19 \\
        \bottomrule
    \end{tabular}
    \label{tab:step_256_12}
\end{table}

\begin{table}[!h]
    \small
    \centering
    \caption{{\bf Cool bits recovery $\boldsymbol{n=256}, \boldsymbol{\log_2 q=20}$} assuming cruel bits have been recovered}
    \begin{tabular}{cc|cccc|cccc|cccc}
        \toprule
         & & \multicolumn{4}{|c}{Linear regression} & \multicolumn{4}{|c}{Stepwise regression} & \multicolumn{4}{|c}{Dual stepwise regression} \\
        Cool bits & Total $h$ & 2M  & 4M & 10M & 20M   & 2M   & 4M & 10M & 20M  & 2M  & 4M & 10M & 20M \\
        \midrule
        20 & 23 & 5 & 14 & 20 & 20 & 5 & 20 & 20 & 20 & 13 & 20 & 20 & 20 \\
        26 & 30 & 2 & 5 & 11 & 13 & 3 & 10 & 19 & 20 & 13 & 17 & 20 & 20 \\
        32 & 37 & 0 & 2 & 4 & 5 & 0 & 4 & 13 & 20 & 5 & 9 & 20 & 20 \\
        38 & 44 & 0 & 0 & 0 & 1 & 0 & 0 & 3 & 11 & 1 & 7 & 13 & 20 \\
        45 & 52 & 0 & 0 & 0 & 0 & 0 & 0 & 0 & 3 & 0 & 2 & 8 & 18 \\
        52 & 60 & 0 & 0 & 0 & 0 & 0 & 0 & 0 & 1 & 0 & 0 & 1 & 14 \\
        \bottomrule
    \end{tabular}
    \label{tab:step_256_20}
\end{table}

\begin{table}[!h]
    \small
    \centering
    \caption{{\bf Cool bits recovery $\boldsymbol{n=512}, \boldsymbol{\log_2 q=28}$} assuming cruel bits have been recovered}
    \begin{tabular}{cc|ccc|ccc|ccc}
        \toprule
         & & \multicolumn{3}{|c}{Linear regression} & \multicolumn{3}{|c}{Stepwise regression} & \multicolumn{3}{|c}{Dual stepwise regression} \\
        Cool bits & Total $h$ & 1M   & 2M  & 4M  & 1M   & 2M   & 4M   & 1M  & 2M  & 4M  \\
        \midrule
        10 & 18 & 20 & 20 & 20 & 20 & 20 & 20 & 20 & 20 & 20 \\
        30 & 53 & 20 & 20 & 20 & 20 & 20 & 20 & 20 & 20 & 20 \\
        40 & 71 & 15 & 20 & 20 & 20 & 20 & 20 & 20 & 20 & 20 \\
        50 & 89 & 5 & 8 & 13 & 14 & 17 & 18 & 17 & 20 & 20 \\
        \bottomrule
    \end{tabular}
    \label{tab:step_512_28}
\end{table}

\begin{table}[!h]
    \small
    \centering
    \caption{{\bf Cool bits recovery $\boldsymbol{n=512}, \boldsymbol{\log_2 q=41}$} assuming cruel bits have been recovered}
    \begin{tabular}{cc|ccc|ccc|ccc}
        \toprule
         & & \multicolumn{3}{|c}{Linear regression} & \multicolumn{3}{|c}{Stepwise regression} & \multicolumn{3}{|c}{Dual stepwise regression} \\
        Cool bits & Total $h$ & 1M   & 2M  & 4M  & 1M   & 2M   & 4M   & 1M  & 2M  & 4M  \\
        \midrule
        40 & 44 & 20 & 20 & 20 & 20 & 20 & 20 & 20 & 20 & 20 \\
        50 & 55 & 20 & 20 & 20 & 20 & 20 & 20 & 20 & 20 & 20 \\
        60 & 66 & 13 & 15 & 18 & 17 & 20 & 20 & 20 & 20 & 20 \\
        70 & 77 & 6 & 12 & 17 & 15 & 18 & 19 & 20 & 20 & 20 \\
        80 & 88 & 0 & 9 & 12 & 10 & 11 & 18 & 19 & 20 & 20 \\
        \bottomrule
    \end{tabular}
    \label{tab:step_512_41}
\end{table}

\section{Ternary secrets}\label{app:ternary}

We report the hardest recovered ternary secret, based on Hamming weight. 

\begin{table}[!h]
\centering
\caption*{\bf Highest Hamming weight and cruel bits recovered - ternary secret.}
\begin{minipage}{0.48\textwidth}
    \small
    \centering
    \caption{\bf $\boldsymbol{n=256}, \boldsymbol{\log_2 q=12}$. }
    \begin{tabular}{l|ccccc}
        \toprule
         & \multicolumn{5}{|c}{Repetition} \\
           & 1x & 2x & 5x & 15x & 50x \\
        \midrule
        \multicolumn{3}{l}{BKZ-reduced data} \\
        4M  & 10/3 & 10/3 & 10/3 & 10/3 & 10/4 \\
        \midrule
        \multicolumn{3}{l}{Synthetic data} \\
        4M  & 10/3 & 10/3 & 10/3 & 10/3 & 10/3 \\
        50M  & 10/4 & 10/4 & 10/4 & 10/4 & - \\
        200M & 10/4 & 10/4 & \textbf{12/5} & - & - \\
        400M & 10/4 & 10/4 & - & - & - \\
       \bottomrule
        \multicolumn{6}{c}{\small Best of 80 models.} \\
    \end{tabular}
    
    \label{tab:results_256_12_ternary}
\end{minipage}
\begin{minipage}{0.48\textwidth}
    \small
    \centering
    \caption{\bf $\boldsymbol{n=512}, \boldsymbol{\log_2 q=28}$}
    \begin{tabular}{l|ccccc}
        \toprule
         & \multicolumn{5}{|c}{Repetition} \\
         & 1x & 2x & 5x & 15x & 50x \\
        \midrule
        \multicolumn{3}{l}{BKZ-reduced data} \\
        4M  & 8/3 & 8/3 & 8/4 & 8/4 & 8/4 \\
        50M  & 8/4 & 10/3 & 10/3 & \textbf{10/4} & - \\
        \midrule
          \multicolumn{3}{l}{Synthetic data} \\
         4M  & 8/3 & 8/3 & 8/3 & 8/3 & 8/4 \\
        50M  & 8/4 & 8/3 & 10/3 & 10/3 & - \\
        200M & 10/3 & 10/3 & \textbf{10/4} & - & - \\
       \bottomrule
        \multicolumn{6}{c}{\small Best of 80 models.} \\
    \end{tabular}
    \label{tab:results_512_28_ternary}
\end{minipage}
\end{table}

\begin{table}[!h]
\centering
\begin{minipage}{0.48\textwidth}
    \small
    \centering
    \caption{\bf $\boldsymbol{n=256}, \boldsymbol{\log_2 q=20}$}
    \begin{tabular}{l|ccccc}
        \toprule
         & \multicolumn{5}{|c}{Repetition} \\
        & 1x & 2x & 5x & 15x & 50x \\
        \midrule
        \multicolumn{3}{l}{BKZ-reduced data} \\
        4M  & 40/5 & 40/5 & 45/5 & 45/5 & 45/5 \\
        50M  & 40/5 & 45/6 & 45/6 & 45/6 & - \\
        200M  & 50/6 & \textbf{55/8} & 55/8 & - & - \\
        400M  & 55/7 & 55/8 & - & - & - \\
        \midrule
          \multicolumn{3}{l}{Synthetic data} \\
        4M  & 40/5 & 40/5 & 40/5 & 45/5 & 45/5 \\
        50M  & 45/5 & 50/6 & 50/6 & 50/6 & - \\
        200M  & 45/5 & 50/6 & 50/6 & - & - \\
        400M  & 50/6 & 50/7 & - & - & - \\
        \bottomrule
        \multicolumn{6}{c}{\small Best of 80 models.} \\
    \end{tabular}
   
    \label{tab:results_256_20_ternary}
\end{minipage}
\begin{minipage}{0.48\textwidth}
    \small
    \centering
    \caption{\bf $\boldsymbol{n=512}, \boldsymbol{\log_2 q=41}$}
    \begin{tabular}{l|cccccc}
        \toprule
         & \multicolumn{5}{|c}{Repetition} \\
         & 1x & 2x & 5x & 15x & 50x \\
        \midrule
        \multicolumn{3}{l}{BKZ-reduced data} \\
        4M  & 60/5 & 60/5 & 60/5 & 65/5 & 65/5 \\
        \midrule
          \multicolumn{3}{l}{Synthetic data} \\
        4M  & 60/5 & 60/5 & 60/5 & 65/5 & 65/5 \\
        50M  & 65/5 & 70/6 & 70/6 & \textbf{75/7} & - \\
        200M  & 70/6 & 70/6 & - & - & - \\
        \bottomrule
        \multicolumn{6}{c}{\small Best of 80 models.} \\
    \end{tabular}
    \label{tab:results_512_41_ternary}
\end{minipage}
\end{table}

\end{document}